# Experimental demonstration of superconducting critical temperature increase in electromagnetic metamaterials

Vera N. Smolyaninova [1) ] , Bradley Yost [1)], Kathryn Zander [1)], M. S. Osofsky [2)], Heungsoo Kim [2)], Shanta Saha [3)], R. L. Greene [3)], and Igor I. Smolyaninov [4)]

[1]*Department of Physics Astronomy and Geosciences, Towson University,*

*8000 York Rd., Towson, MD 21252, USA*

[2] *Naval Research Laboratory, Washington, DC 20375, USA*

[3]*Department of Physics, Center for Nanophysics & Advanced Materials, University of Maryland, College Park, MD 20742, USA*

[4]*Department of Electrical and Computer Engineering, University of Maryland, College Park, MD 20742, USA*

Correspondence to smoly@umd.edu

**A recent proposal that the metamaterial approach to dielectric response engineering may increase the critical temperature of a composite superconductor-dielectric metamaterial has been tested in experiments with compressed mixtures of tin and barium titanate nanoparticles of varying composition. An increase of the critical temperature of the order of $\Delta T$ ~ 0.15 K compared to bulk tin has been observed for 40% volume fraction of barium titanate nanoparticles. Similar results were also obtained with compressed mixtures of tin and strontium titanate nanoparticles.**

Metamaterials are artificial materials built from conventional microscopic materials in order to engineer their properties in a desired way. Such artificial metamaterials offer novel ways for controlling electromagnetic [1], acoustic [2], elastic [3], and thermal [4] properties of matter. Very recently, it was proposed that the metamaterial approach to dielectric response engineering might also increase the critical temperature of a composite superconductor-dielectric metamaterial [5,6]. Indeed, electromagnetic properties are known to play a very important role in the pairing mechanism and charge dynamics of high $T_c$ superconductors [7]. Moreover, shortly after the original work by Bardeen, Cooper and Schrieffer (BCS) [8], Kirzhnits *et al.* formulated a complementary description of superconductivity in terms of the dielectric response function of the superconductor $\varepsilon_{eff}(q,\omega)$, which was treated as a homogeneous medium [9]. In this description the phonon-mediated electron-electron interaction in superconductors is expressed in the form of the effective Coulomb potential

$$V(\vec{q},\omega)=\frac{4\pi e^2}{q^2\varepsilon_{eff}(\vec{q},\omega)}, \qquad (1)$$

where $\varepsilon_{eff}(q,\omega)$ is the linear dielectric response function of the superconductor treated as a homogeneous effective medium. While the thermodynamic stability condition implies that $\varepsilon_{eff}(q,0)>0$, the dielectric response function at higher frequencies may become large and negative, which accounts for the weak net attraction and pairing of electrons in the superconducting condensate. Based on this approach, Kirzhnits et al. derived explicit expressions for the superconducting gap $\Delta$ and critical temperature, $T_c$, of the superconducting transition as a function of the dielectric response function, and demonstrated that their expressions agree with the BCS result.

The key observation made in Refs. [5,6] was that the "homogeneous medium" approximation may remain valid even if the basic structural elements of the material are not simple atoms or molecules. It was suggested that artificial "metamaterials" may be created from much bigger building blocks, and the electromagnetic properties of these fundamental building blocks may be engineered in such a way that the attractive pairing interaction described by Eq. (1) is maximized. The two potential solutions described in [5,6] require minimization of $\varepsilon_{eff}(q,\omega)$ within a substantial portion of the relevant four-momentum spectrum ($|\vec{q}| \le 2k_F$, $\omega \le$ BCS cutoff around the Debye energy), while keeping the real part of $\varepsilon_{eff}(q,\omega)$ negative. These solutions involve either the epsilon-near-zero (ENZ) [10], or hyperbolic metamaterial [11] scenarios, which naturally lead to broadband small and negative $\varepsilon_{eff}(q,\omega)$ behavior in substantial portions of the relevant four-momentum spectrum. In the isotropic ENZ scenario the effective pairing interaction is described by Eq. (1), where $\varepsilon_{eff}(q,\omega)$ is assumed to characterize the macroscopic electrodynamic properties of the composite metamaterial. On the other hand, in the hyperbolic metamaterial scenario the effective Coulomb potential from Eq. (1) assumes the form

$$V(\vec{q},\omega) = \frac{4\pi e^2}{q_z^2\varepsilon_2(\vec{q},\omega) + \left(q_x^2 + q_y^2\right)\varepsilon_1(\vec{q},\omega)} \quad , \tag{2}$$

where $\varepsilon_{xx} = \varepsilon_{yy} = \varepsilon_1$ and $\varepsilon_{zz} = \varepsilon_2$ have opposite signs [5,6]. As a result, the effective Coulomb interaction of two electrons may become attractive and very strong along spatial directions where

$$q_z^2\varepsilon_2(\vec{q},\omega) + \left(q_x^2 + q_y^2\right)\varepsilon_1(\vec{q},\omega) \approx 0 \tag{3}$$

and negative, which indicates that the superconducting order parameter must be strongly anisotropic. This situation resembles the hyperbolic character of such high $T_c$ superconductors as BSCCO [5,6]. However, engineering such ENZ or hyperbolic metamaterials constitutes a much more challenging task compared to typical applications of superconducting metamaterials suggested so far [12], which only deal with metamaterial engineering on scales which are much smaller than the microwave or RF wavelengths. Since the superconducting coherence length (the size of the Cooper pair) is $\xi$~100 nm in a typical BCS superconductor, the metamaterial unit size must fall within the window between ~0.3 nm (given by the atomic scale) and $\xi$~100 nm scale of a typical Cooper pair. Moreover, the coherence length of the metamaterial superconductor must be determined in a self-consistent manner. The coherence length will decrease with increasing $T_c$ of the metamaterial superconductor since the approach of Kirzhnits et al. gives rise to the same BCS-like relationship between the superconducting gap $\Delta$ and the coherence length $\xi$ [9]:

$$\left(\frac{\xi}{V_F}\right)\Delta \sim \hbar\,, \tag{4}$$

where $V_F$ is the Fermi velocity. Therefore, the metamaterial structural parameter (such as the inter-layer or inter-particle distance, etc.) which must remain smaller than the coherence length, defines the limit of critical temperature enhancement.

Let us demonstrate that tuning electron-electron interaction is indeed possible using the epsilon-near-zero (ENZ) metamaterial approach, which is based on intermixing metal and dielectric components in the right proportions. A negative $\varepsilon \approx 0$ ENZ metamaterial would maximize attractive electron-electron interaction given by Eq. (1). We consider a random mixture of a superconducting “matrix” with dielectric "inclusions" described by the dielectric constants $\varepsilon_m$ and $\varepsilon_i$, respectively in the frequency range of interest. In the Maxwell-Garnett approximation the effective dielectric constant, $\varepsilon_{eff}$, of the metamaterial may be obtained as

$$\left(\frac{\varepsilon_{eff}-\varepsilon_m}{\varepsilon_{eff}+2\varepsilon_m}\right)=\delta_i\left(\frac{\varepsilon_i-\varepsilon_m}{\varepsilon_i+2\varepsilon_m}\right), \tag{5}$$

where $\delta_i$ is the volume fraction of the inclusions (considered to be small) [13]. The explicit expression for $\varepsilon_{eff}$ is then written as

$$\varepsilon_{eff}=\varepsilon_m\frac{(2\varepsilon_m+\varepsilon_i)-2\delta_i(\varepsilon_m-\varepsilon_i)}{(2\varepsilon_m+\varepsilon_i)+\delta_i(\varepsilon_m-\varepsilon_i)} \tag{6}$$

The ENZ condition ($\varepsilon_{eff}\approx 0$) is then obtained for

$$\delta_i=\frac{2\varepsilon_m+\varepsilon_i}{2(\varepsilon_m-\varepsilon_i)}, \tag{7}$$

which means that $\varepsilon_m$ and $\varepsilon_i$ must have opposite signs, and $\varepsilon_i\approx -2\varepsilon_m$ so that $\delta_i$ will be small. This consideration indicates that the attractive electron-electron interaction in a superconducting metamaterial may indeed be increased by using the correct amount of dielectric. However, $\varepsilon_i$ of the dielectric must be very large, since $\varepsilon_m$ of the metal component, typically given by the Drude model in the far infrared and THz ranges

$$\varepsilon_m=\varepsilon_{m\infty}-\frac{\omega_p^2}{\omega(\omega+i\gamma)}\approx-\frac{\omega_p^2}{\omega(\omega+i\gamma)}, \tag{8}$$

is large and negative (where $\varepsilon_{m\infty}$ is the dielectric permittivity of metal above the plasma edge, $\omega_p$ is its plasma frequency, and $\gamma$ is the inverse free propagation time). Ferroelectric materials having large positive $\varepsilon_i$ in the relevant frequency range should be a very good choice for use as the dielectric. Moreover, if the high frequency behavior of $\varepsilon_i$ may be assumed to follow the Debye model [14]:

$$\mathrm{Re}\,\varepsilon_i = \frac{\varepsilon_i(0)}{1+\omega^2\tau^2} \approx \frac{\varepsilon_i(0)}{\omega^2\tau^2}, \tag{9}$$

then broadband ENZ behavior results due to the similar $\sim\omega^{-2}$ functional behavior of $\varepsilon_i$ and $\varepsilon_m$ in the THz range (compare Eqs. (8) and (9)). Thus, the BCS theory assumption of a constant attractive interaction from low frequencies (of the order of the gap energy) to the range of the BCS cutoff around the Debye energy would be approximately satisfied. However, this is not a strict requirement, and the attractive interaction in metamaterial superconductors described by eqs.(1) may depend on frequency due to dispersive behavior of $\varepsilon_{eff}(q,\omega)$.

We should also point out that the Maxwell-Garnett based analysis of Eqs. (1) and (5) indicates that even far from ENZ conditions a superconductor-dielectric metamaterial must have larger $\Delta$ and higher $T_c$ compared to the original undiluted superconducting host. Indeed, even in the limit $\varepsilon_i << -\varepsilon_m$ and small $\delta_i$ the Maxwell-Garnett approximation (Eq. (5)) results in a decrease of $\varepsilon_{eff}$

$$\varepsilon_{eff} = \varepsilon_m \frac{(2\varepsilon_m+\varepsilon_i)-2\delta_i(\varepsilon_m-\varepsilon_i)}{(2\varepsilon_m+\varepsilon_i)+\delta_i(\varepsilon_m-\varepsilon_i)} \approx \varepsilon_m \frac{2\varepsilon_m-2\delta_i\varepsilon_m}{2\varepsilon_m+\delta_i\varepsilon_m} \approx \varepsilon_m\left(1-\frac{3}{2}\delta_i\right) \tag{10}$$

producing an increase in the potential:

$$V(\vec{q},\omega) = \frac{4\pi e^2}{q^2\varepsilon_{eff}(\vec{q},\omega)} \approx \frac{4\pi e^2}{q^2\varepsilon_m(\vec{q},\omega)}(1+\frac{3}{2}\delta_i) \tag{11}$$

which should be proportional to the increase in $\Delta$ and $T_c$, and which in this limit does not depend on the particular choice of dielectric $\varepsilon_i$. According to Eq. (4), compared to the parent pure superconductor, the metamaterial coherence length $\xi_{MM}$ will decrease as

$$\xi_{MM} = \frac{\xi_0}{\left(1+\frac{3}{2}\delta_i\right)} \tag{12}$$

where $\xi_0$ is coherence length of the parent superconductor. Therefore, the metamaterial design should make sure that the typical metamaterial structural parameter remains much below $\xi_{MM}$. This result constitutes a proof of principle of the metamaterial approach and can be experimentally tested. While ideal ENZ conditions must lead to large increases in $T_c$, detailed calculations of the enhancement in real materials are much more difficult, since they require detailed knowledge of $\varepsilon_m(q, \omega)$ and $\varepsilon_i(q, \omega)$ of the metamaterial components. On the other hand, evaluation of the maximum critical temperature $T_c^{max}$ of the superconducting transition from the point of view of the electromagnetic approach performed by Kirzhnits *et al.* in Ref. [9] produced a very optimistic $T_c^{max}$~300 K estimate at $E_F$ ~10 eV. According to Eq. (4), the corresponding $\xi_{MM}$ ~ 48 nm leaves substantial room for metamaterial engineering.

Following the theoretical motivation described above, we have fabricated several composite superconductor-dielectric metamaterials of varying composition and measured the predicted increase in the superconducting critical temperature compared to the bulk parent superconductor. Our superconducting metamaterial samples were prepared using commercially available tin and barium titanate nanoparticles obtained from the US Research Nanomaterials, Inc. The nominal diameter of the $BaTiO_3$ nanoparticles was 50 nm, while tin nanoparticle size was specified as 60-80 nm. Both nanoparticle sizes fall substantially below the superconducting coherence length in pure tin $\xi_{Sn}$ ~230 nm. Our choice of materials was based on results of numerical calculations of the real part of the dielectric constant of the $Sn/BaTiO_3$ mixture shown in Fig.1. These calculations were based on the measured dielectric properties of Sn [15] and $BaTiO_3$ [16] extrapolated by Eqs. (8) and (9), respectively, and on the Maxwell-Garnett expression, Eq. (6), for the dielectric permittivity of the mixture. These calculations indicated that it was possible to achieve broadband ENZ conditions in the 30-50% range of the volume fraction of $BaTiO_3$ in the frequency range of relevance around $h\nu \sim kT_c$ of pure tin.

The $Sn/BaTiO_3$ superconducting metamaterials were fabricated by combining the given amounts of Sn and $BaTiO_3$ nanoparticle powders by volume into a single test tube filled with de-ionized water. The resulting suspensions were sonicated and magnetically stirred for 30 minutes, followed by water evaporation. The remaining mixtures of Sn and $BaTiO_3$ nanoparticles were compressed into test pellets using a hydraulic press. A typical Scanning Electron Microscopy (SEM) image of the resulting $Sn/BaTiO_3$ composite metamaterial is shown in Fig. 2. The original compressed nanoparticles are clearly visible in the image. The resulting average metamaterial composition has been verified after fabrication using an SEM with energy dispersive X-ray spectroscopy (EDS). The compositional analysis of an area of the samples of about 1 μm in diameter is consistent with this nominal composition. Such EDX spectra were used to establish the homogeneous character of the fabricated composite metamaterials.

The superconducting critical temperature $T_c$ of various $Sn/BaTiO_3$ metamaterials was measured via the onset of diamagnetism for samples with different volume fractions of $BaTiO_3$ using a MPMS SQUID magnetometer. The zero field cooled (ZFC) magnetization per unit mass for several samples with varying concentrations of $BaTiO_3$ is plotted in Fig. 3(a). The temperatures of the onset of the superconducting transition and the temperatures of the midpoint of the transition are plotted in Fig. 4. The $T_c$ increased from the pure Sn value of 3.68 K with increasing $BaTiO_3$ concentration to a maximum $\Delta T_c \sim 0.15$ K or 4% compared to the pure tin sample for the 40% sample followed by $T_c$ decreasing at higher volume fractions. A pure tin sample was prepared from pressed tin nanoparticles of the same diameter using the same method of preparation. The value of $T_c$ agreed with expected value of pure Sn. The magnetization of the composite samples in the superconducting state is comparable in magnitude with the pure tin sample. The increase of $T_c$ and its dependence on effective dielectric constant determined by the concentration of $BaTiO_3$ agrees qualitatively with the Maxwell-Garnett theory-based calculations described above (see Fig. 1). Moreover, decrease of $T_c$ at higher volume fraction of $BaTiO_3$, which is apparent from Fig. 4(b) also agrees well with the Maxwell-Garnett theory, since $\varepsilon_{eff}(q,\omega)$ changes sign to positive for higher $BaTiO_3$ concentrations.

In order to verify the reproducibility of these results, our measurements were repeated for several sets of nanocomposite samples (compare 40% $BaTiO_3$volume fraction samples in Fig. 3(a) and Fig. 3(c)). Moreover, to further verify our understanding of the metamaterial mechanism of $T_c$ increase we have fabricated a set of samples using either strontium titanate (Fig. 3(b)) or diamond (Fig. 3(c)) nanoparticles (instead of barium titanate) using the same fabrication technique described above. Similar to $BaTiO_3$, strontium titanate also has large positive $\varepsilon$ in the relevant frequency range. The superconducting critical temperature $T_c$ of the Sn/$SrTiO_3$ metamaterial samples was measured for samples with different volume fractions of $SrTiO_3$ nanoparticles as shown in Fig. 3(b) producing volume fraction dependencies, which are similar to those obtained for Sn/$BaTiO_3$ samples. On the other hand, diamond has rather low dielectric constant $\varepsilon_{dia}$~5.6, which stays almost constant from the visible to RF frequency range. As demonstrated in Fig. 3(c) and 4(b), in agreement with our Maxwell-Garnett based analysis (Eqs. (5-11)), tin/diamond nanoparticle metamaterial samples do not exhibit much increase in $T_c$ compared to the pure tin samples.

Thus, we have established that the metamaterial approach to dielectric response engineering can increase the critical temperature of a composite superconductor-dielectric metamaterial. We expect that these initial results may be further improved by refining the metamaterial structural parameter, which ideally must fall well below the superconducting coherence length. Moreover, our superconducting metamaterial concept is not limited to simple geometries made of metal-dielectric nanoparticles or nanolayer composites. It can be further refined by taking full advantage of metamaterial electromagnetic theory and 3D nanoengineering. Such a refinement may lead to a considerable increase of the effective electron-electron pairing interaction described by Eq. (1). Since the metamaterial dimensions required for engineering the electron-electron interaction approach the nanometer scale, another potentially important issue is the applicability of describing nanoscale metal and dielectric layers using their macroscopic dielectric constants. This issue is well known and extensively studied in the fields of nanophotonics and electromagnetic metamaterials. The electromagnetic response of thin metal layers is indeed known to exhibit a weak oscillatory dependence on layer thickness due to quantum mechanical effects, such as the formation of electron standing waves inside the thin layer [17]. While this effect indeed influences the effective dielectric constant of a metal layer, for all practical purposes this is a weak effect. On the other hand, the dielectric constant of dielectric materials does not depend on layer dimensions until the atomic scale is reached. This fact has been verified in experiments on surface plasmon resonance [18]. Another limit on the smallest nanoparticle size $a \sim \hbar / \sqrt{2mkT_c}$ ~10 nm is defined by the superconducting size effect: when spacing between the electron levels in tin nanoparticles becomes larger than the superconducting energy gap, the superconductivity cannot exist anymore. However, this already small limit reduces even further with the increase in superconducting gap.

In conclusion, we have demonstrated that recent developments in the field of electromagnetic metamaterials, such as the development of epsilon near zero (ENZ) and hyperbolic metamaterials, may be used to engineer the dielectric response of composite superconducting metamaterials on the sub-100 nm scale in order to increase the superconducting critical temperature of the metamaterial. Our proposal has been tested in experiments with metamaterials based on compressed mixtures of tin and barium titanate nanoparticles of varying composition. An increase of the critical temperature of the order of $\Delta T \sim 0.15$ K compared to bulk tin has been observed for ~ 40% volume fraction of barium titanate nanoparticles.

This work was supported in part by NSF grant DMR-1104676 at Towson and AFOSR grant FA9550-09-1-0603 at Maryland. We are grateful to S. Anlage for fruitful discussion, and to M. Monk for experimental help.

I.S. and V.S. wrote the main manuscript text, V.S., B.Y. and K.Z. fabricated samples, V.S., M.O., H.K., S.S., and R.G. collected experimental data. I.S. and V.S. developed the theory of metamaterial superconductivity. All authors reviewed the manuscript.

**Additional Information**

Competing financial interests.

The authors declare no competing financial interests.

**Figure Captions**

**Figure 1.** Numerical calculations of the real part of the dielectric constant of the Sn/$BaTiO_3$ mixture as a function of volume fraction of $BaTiO_3$ performed with 10% steps.

**Figure 2.** (a) SEM image of the composite Sn/$BaTiO_3$ metamaterial with 30% volume fraction of $BaTiO_3$ nanoparticles. Individual compressed nanoparticles are clearly visible in the image. (b) Schematic diagram of the metamaterial sample geometry.

**Figure 3**. Temperature dependence of zero field cooled magnetization per unit mass for several samples with varying concentration of (a) $BaTiO_3$ and (b) $SrTiO_3$ measured in magnetic field of 10 G. (c) Comparison of zero field cooled magnetization measurements performed for pure tin nanoparticle samples, and samples containing 40% volume fraction of either $BaTiO_3$ or diamond nanoparticles.

**Figure 4.** (a) The temperatures of the onset of the superconducting transition (blue squares) and the temperatures of the midpoint of the transition (red circles) plotted as a function of volume fraction of $BaTiO_3$. Line is a guide to the eye. (b) Comparison of the transition midpoint plotted as a function of volume fraction of $BaTiO_3$ (red circles) and diamond (blue circles).

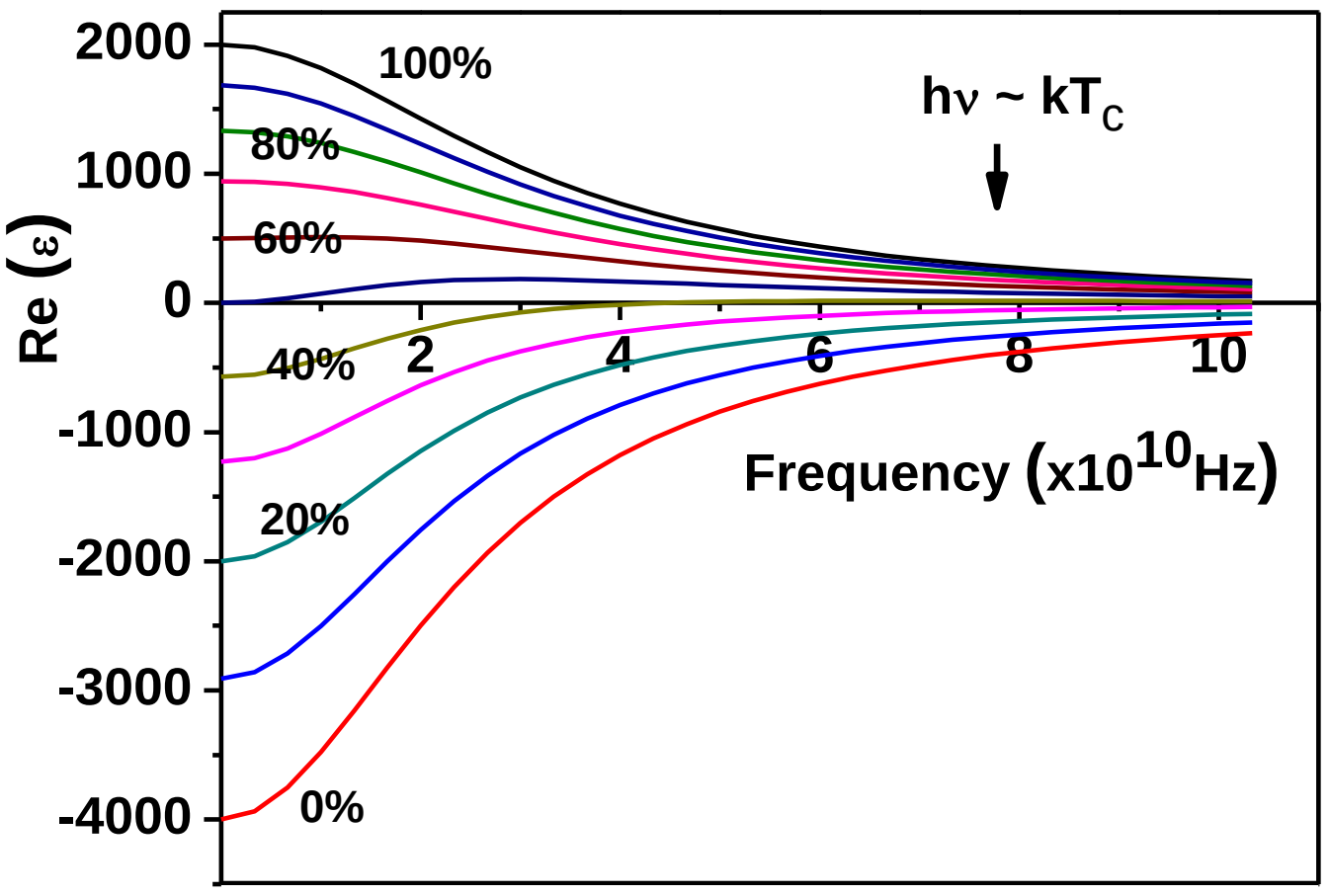

Fig.1

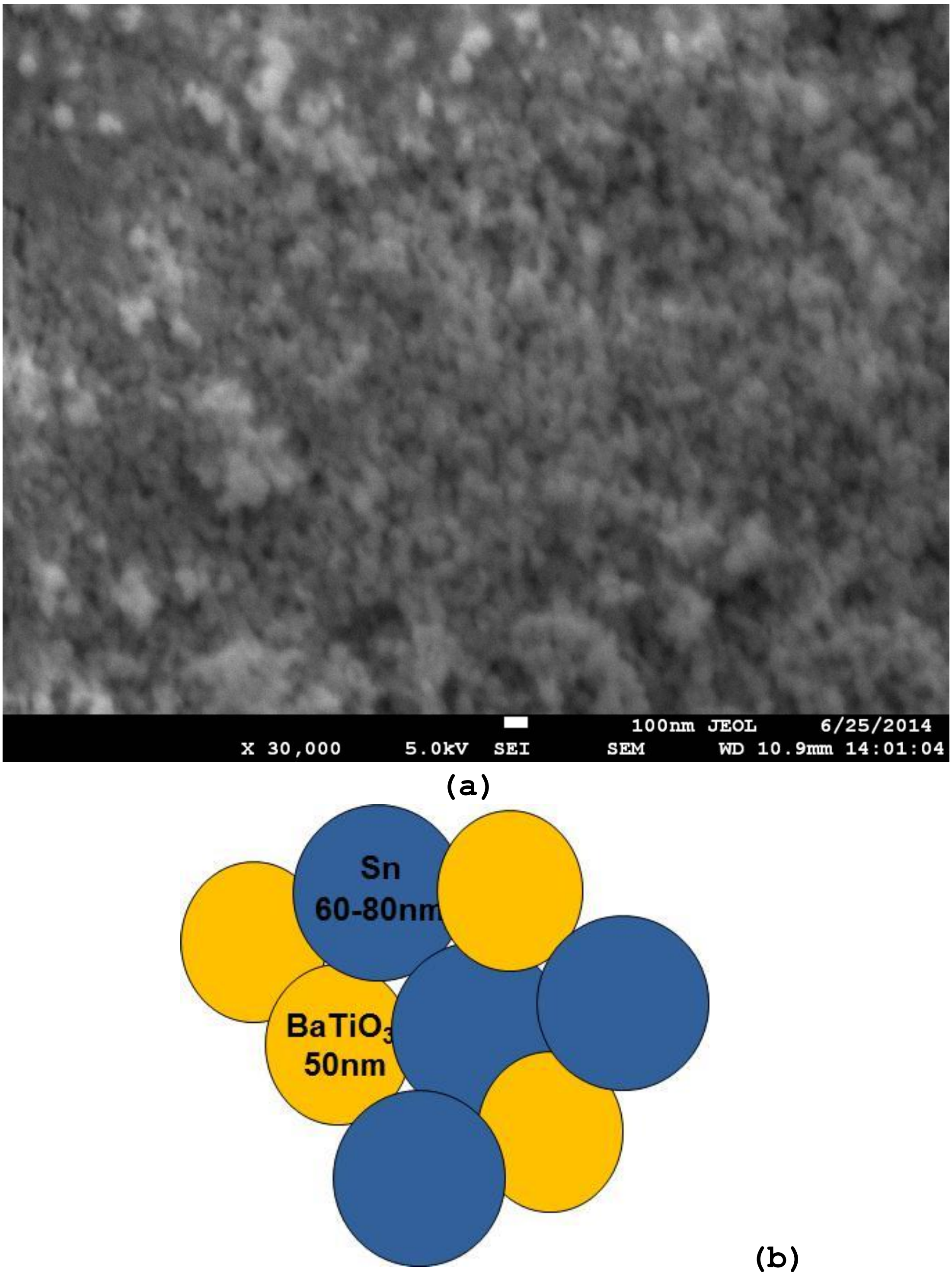
100nm JEOL 6/25/2014
X 30,000 5.0kV SEI SEM WD 10.9mm 14:01:04
(a)
Sn
60-80nm
BaTiO3
50nm
(b)

Fig.2

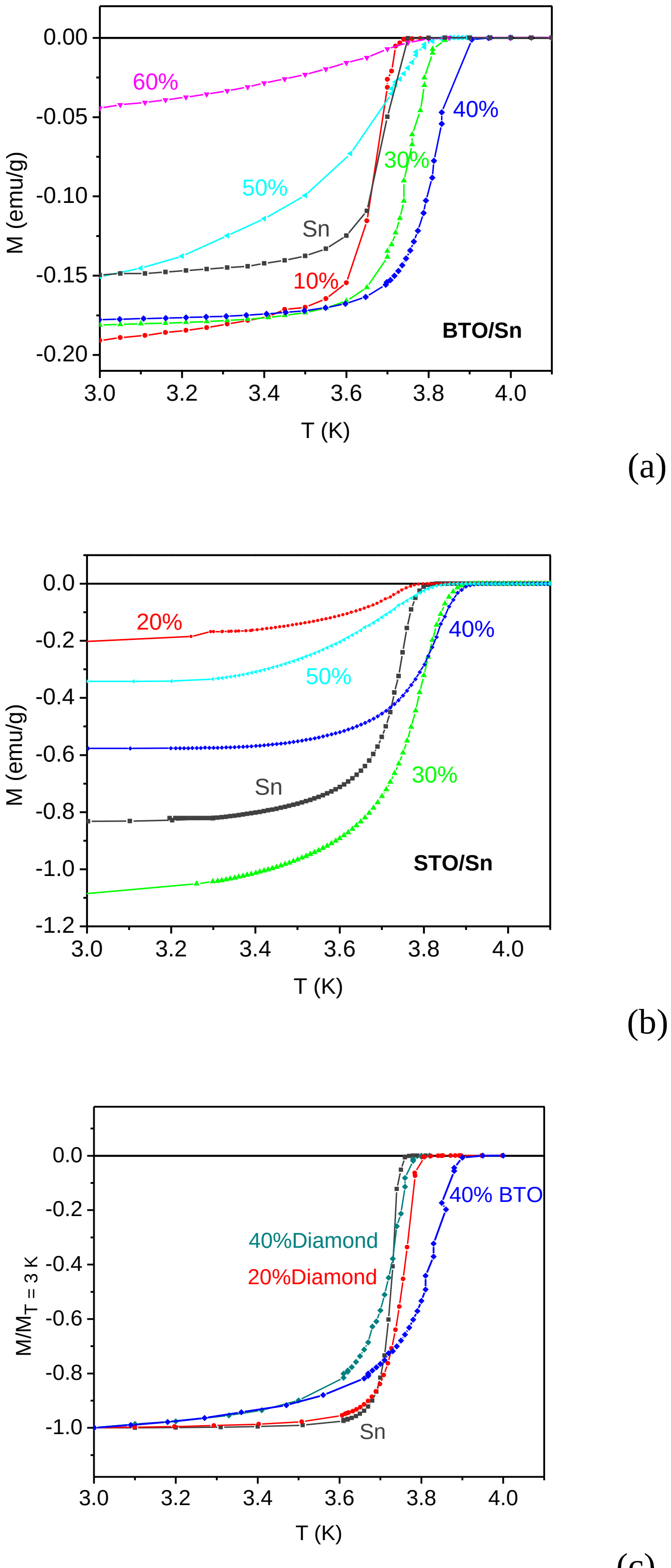

Fig. 3

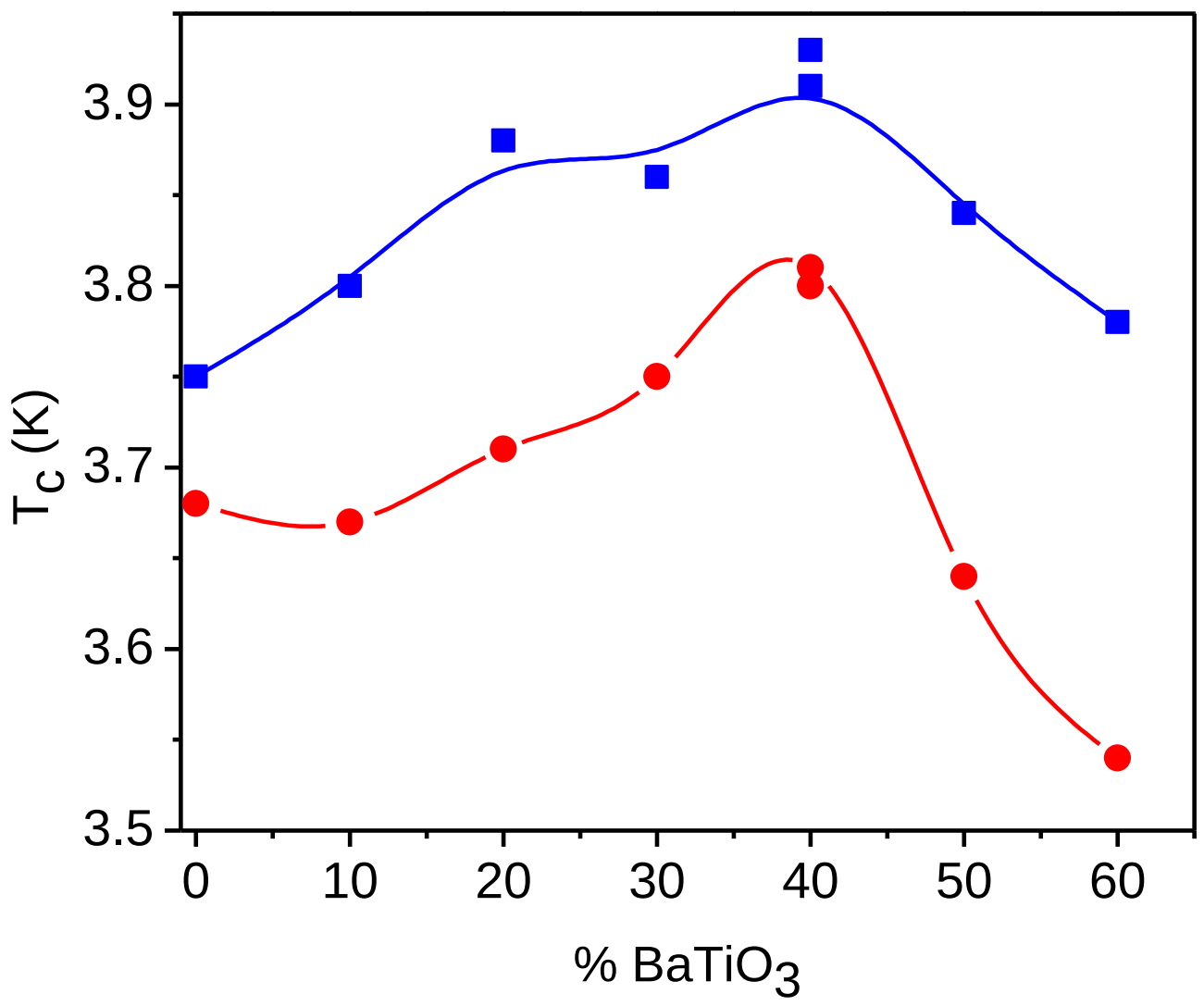

(a)

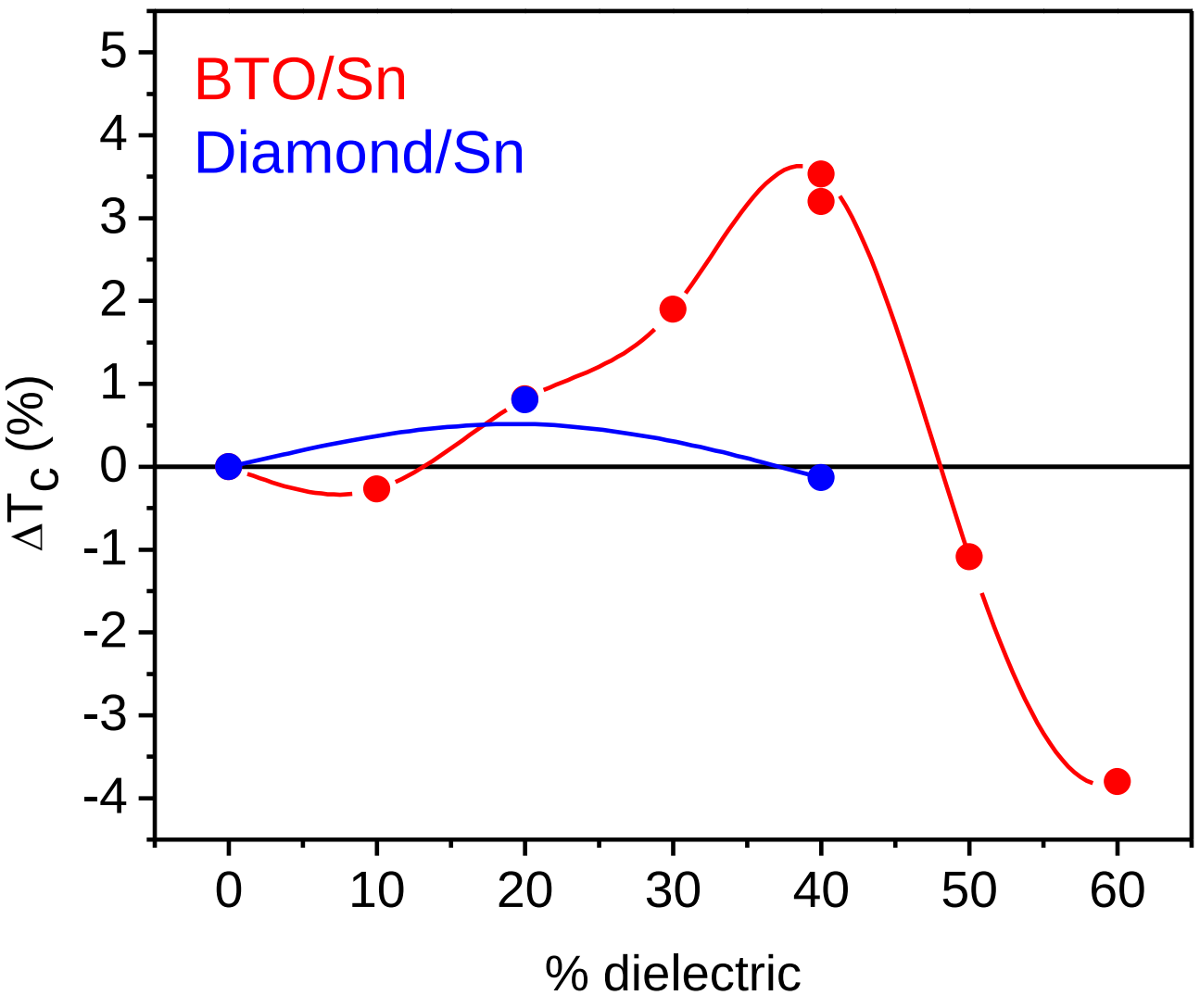

(b)

Fig. 4